\documentclass[11pt]{scrartcl} 
\ifx\pdfminorversion\undefined
\else
\fi

\usepackage{fullpage,epsfig}
\usepackage{comment}
\usepackage{xcolor}
\usepackage{longtable}
\usepackage{multirow,array}
\usepackage{authblk}
\usepackage{parskip}
\PassOptionsToPackage{hyphens}{url}
\usepackage{hyperref}
\usepackage{color,xspace}
\usepackage{cancel}

\usepackage{graphicx}
\usepackage{caption}
\usepackage{subcaption}

\usepackage{enumitem}
\usepackage{amsmath}
\usepackage{listings}
\lstdefinestyle{PythonStyle}{
    language=Python,
    basicstyle=\ttfamily\footnotesize, 
    keywordstyle=\color{blue}\bfseries,
    identifierstyle=\color{black},
    stringstyle=\color{orange},
    commentstyle=\color[rgb]{0.13,0.54,0.13}, 
    showstringspaces=false,
    columns=fullflexible, 
    breaklines=true,
    frame=single, 
    frameround=tttt,
    rulesepcolor=\color{gray},
    numbers=left, 
    numberstyle=\tiny\color{gray},
    stepnumber=1,
    tabsize=4,
    captionpos=b, 
    backgroundcolor=\color{gray!10}, 
    literate={å}{{\aa}}1
             {ø}{{\o}}1    
}

\usepackage[utf8]{inputenc}

\newcommand{\remove}[1]{}

\title{Cenergy3: An API for City Energy 3D Modeling}

\author{Shiliang Zhang$^*$\qquad Sabita Maharjan\thanks{Shiliang Zhang and Sabita Maharjan are with Department of Informatics, University of Oslo, Norway (e-mail: \{shilianz, sabita\}@uio.no).}}

\begin{document}
\maketitle

\begin{abstract}
The efficient management and planning of urban energy systems require integrated three-dimensional (3D) models that accurately represent both consumption nodes and distribution networks. This paper introduces our developed geospatial Application Programming Interface (API) that automates the generation of 3D urban digital model from open data. The API synthesizes data from OpenTopography, OpenStreetMap, and Overture Maps in generating 3D models. The rendered model visualizes and contextualizes power grid infrastructure alongside the built environment and transportation networks. The API provides interactive figures for the 3D models, which are essential for analyzing infrastructure alignment and spatially linking energy demand nodes (buildings) with energy supply (utility grids). Our API leverages standard \texttt{Web Mercator} coordinates (\texttt{EPSG:3857}) and \texttt{JSON} serialization to ensure interoperability within smart city and energy simulation platforms.
\end{abstract}

\begin{figure}[hbtp]
    \centering
    \includegraphics[width=0.65\linewidth]{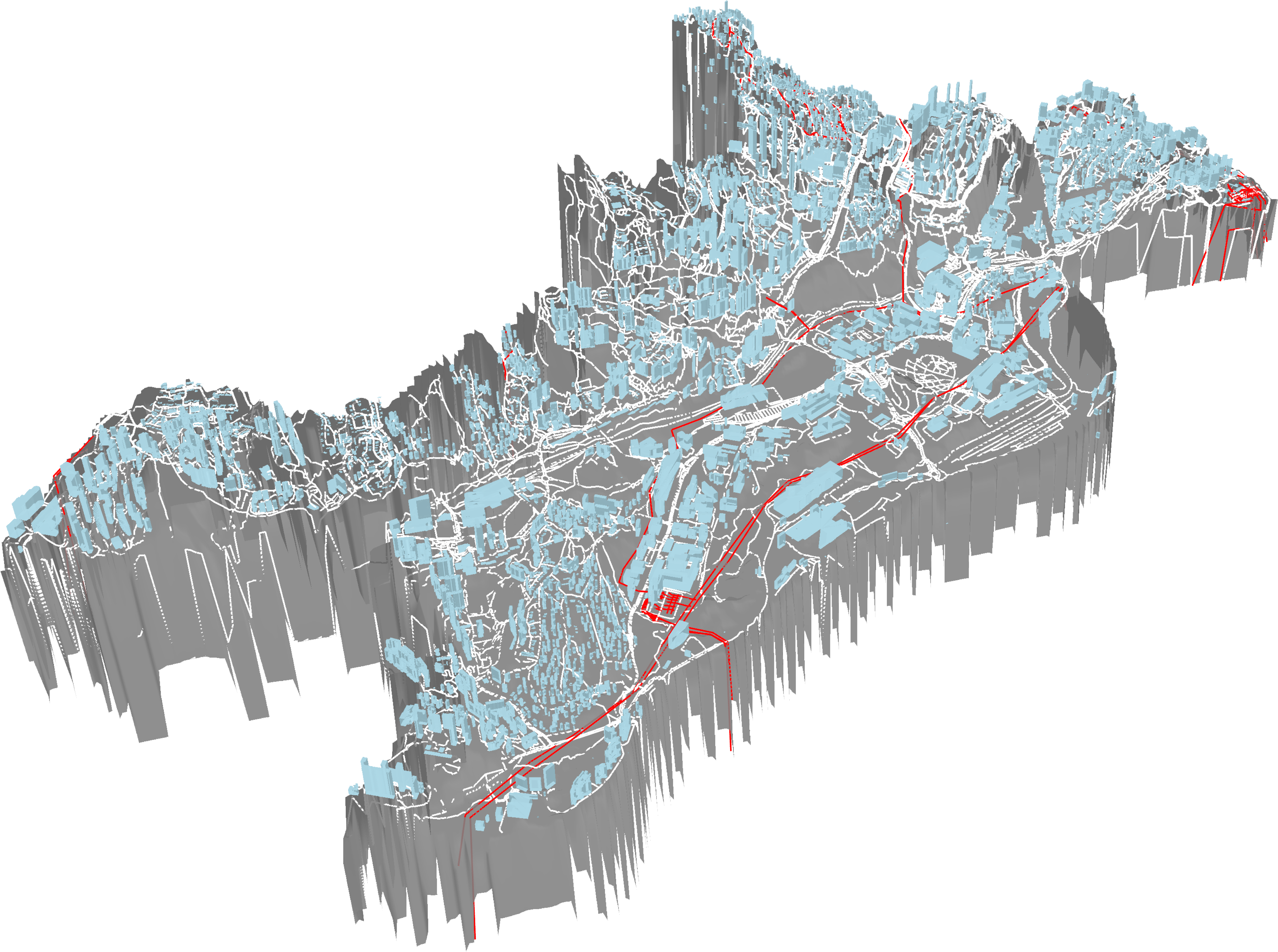}
    \caption{An example of the resulted 3D model for Alna, Oslo, Norway. The white lines are road networks, the red lines are utilty grids, and the light blue blocks represent buildings.}
    \label{fig:demonstration}
\end{figure}

\section{Introduction}

The contemporary urban environment is characterized by an exponential increase in accumulated geospatial datasets relevant to energy systems, including consumption, production, energy price and infrastructure, \textit{etc}~\cite{DBLP:journals/corr/abs-2510-09698,hersbach2019era5,11174415,weinand2019spatial,9575223,DBLP:journals/corr/abs-2203-08630}. Analyzing this information is critical for drawing insights and generating new knowledge that can promote objectives such as energy infrastructure planning, energy resilience towards penetrating renewables, optimizing power flow, and robust demand-supply analyses, prediction and prognosis, \textit{etc}~\cite{10538433,DBLP:journals/corr/abs-2509-17095,oskar2025exploring,10738058}.

Nevertheless, the direct use of high-resolution energy data, \textit{e.g.}, granular smart meter readings that contain rich details about energy activities, presents privacy challenges~\cite{10738107}. Such energy data can be sensitive and potentially expose individual occupancy patterns, lifestyle details, and personal routines~\cite{zhang2020privacy,DBLP:journals/sensors/ZhangHSSA23}, thus compromising privacy~\cite{wylde2022cybersecurity,DBLP:journals/corr/abs-2503-03539,DBLP:journals/corr/abs-2312-11564}.

Open geospatial energy data offers a better, privacy-preserving alternative for large-scale energy analysis and planning~\cite{SALVALAI2024114500,el2024comprehensive}. Data source like OpenStreetMap~\cite{9119753} provides high-fidelity, anonymized proxies for energy-relevant features—such as building footprints, road access, and physical infrastructure location—without requiring access to sensitive and individual details.

However, a gap exists in translating disparate and data-rich analyses into actionable, intuitive insights. There is a lack of effective tools for visualizing the spatial context of energy infrastructures within the city~~\cite{DBLP:journals/corr/abs-2510-09698}. Static reports fail to capture the relationships between terrain, buildings, and the physical alignment of power distribution networks. This deficiency in contextual visualization hinders effective communication among planners, engineers, and policymakers.

This paper presents an API that is free and designed to address this gap by generating geospatially-aware 3D model for any specific area. The API automates the creation of a 3D model that synthesizes terrain data from OpenTopography, detailed building semantics (especially height) from Overture Maps, and infrastructure topology (roads and power grids) from OpenStreetMap. Fig.~\ref{fig:demonstration} shows an example of the 3D model generation. By visualizing the power grid infrastructure alongside the built environment and transportation networks, our API provides the essential tool needed to spatially link energy data with the physical city, offering a novel solution for advancing urban energy analysis and planning.

\section{Cenergy3 API: The mechanisms behind}

The developed Cenergy3 API relies entirely on public, open data sources. The API allows client applications to request 3D models via a simple HTTP request. The backend processing pipeline consists of four distinct stages: Geocoding, terrain Processing, vector draping, and building extrusion. We elaborate those stages below.

\subsection{Geocoding}
Upon receiving a request, the API utilizes \texttt{OSMnx} to geocode the target place name into a georeferenced polygon boundary. A bounding box is extracted from this polygon to define the area of interest. This bounding box coordinates data retrieval to ensure spatial alignment.

\subsection{Terrain processing}
\begin{figure}[tbhp]
    \centering
    \includegraphics[width=0.5\linewidth]{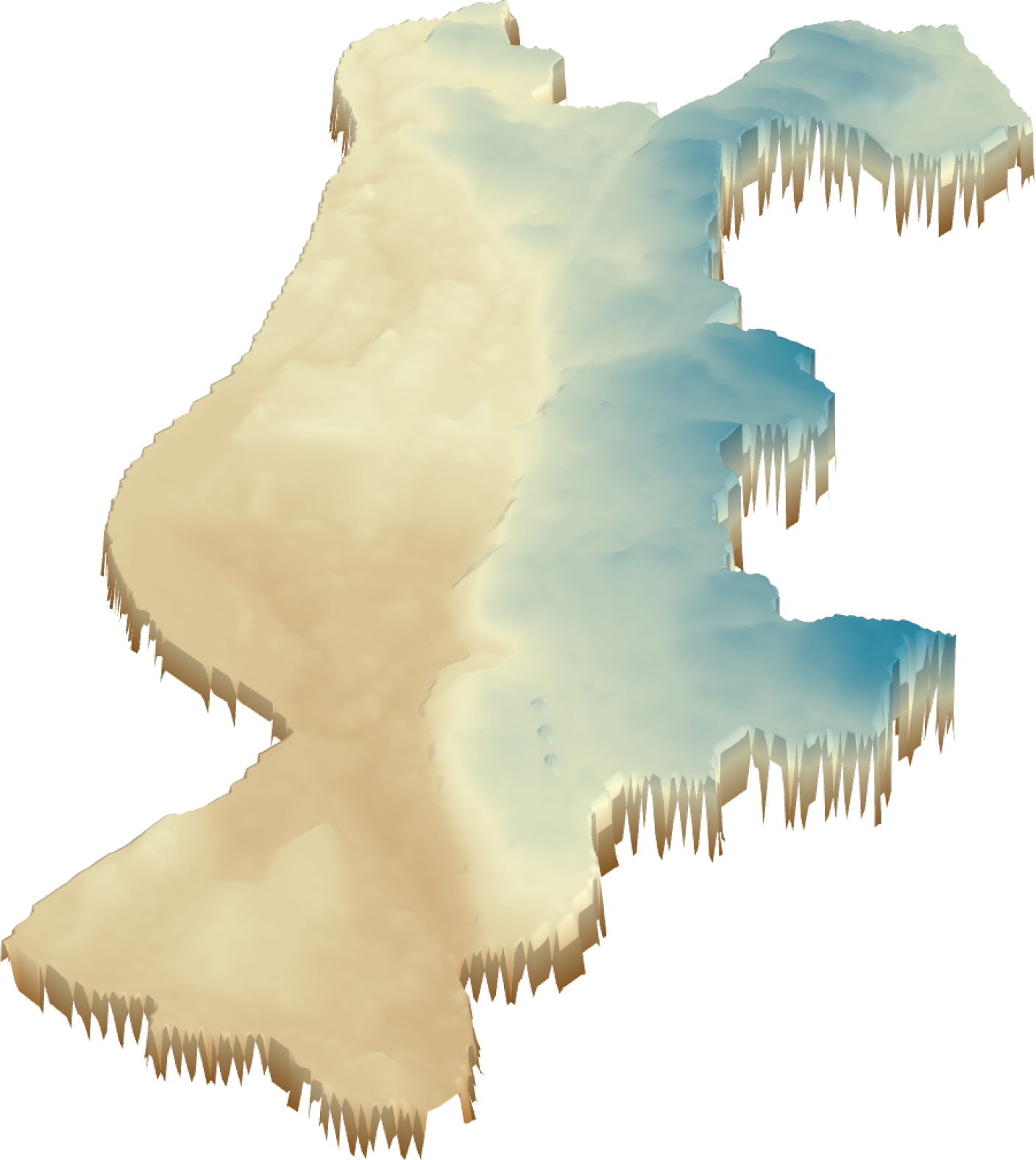}
    \caption{An example of the visualized terrain (for the area of Alna, Oslo, Norway).}
    \label{fig:terrain}
\end{figure}

The digital elevation model (DEM) is the foundational layer of the 3D modeling. We fetch DEM information represented by global copernicus 30m (COP30) DEM data via the OpenTopography. An additional API is needed to access the service from OpenTopography, which is free from the OpenTopography platform. Then we reproject the COP30 DEM data from its native coordinating reference system (EPSG:4326) to the Web Mercator projection (EPSG:3857). This ensures metric accuracy for 3D rendering. The reprojected information is then converted into a triangular mesh to generate the terrain of the considered area. We provide an example of the terrain processed in Fig.~\ref{fig:terrain}. Particularly, we generate vertices of the terrain for every valid pixel in the COP30 DEM data. We construct the faces of the terrain by triangulating adjacent pixel centers. This standardizes the geometry of the elevation data before rendering a 3D model.

\subsection{Vector data draping}
A key challenge in mixing 2D vector data of roads and power lines with 3D terrain is the vertical alignment. In addressing this, the API implements a draping mechanism illustrated below.

\begin{figure}[tbhp]
    \centering
    \begin{subfigure}[b]{0.45\textwidth}
        \centering
        \includegraphics[width=\linewidth]{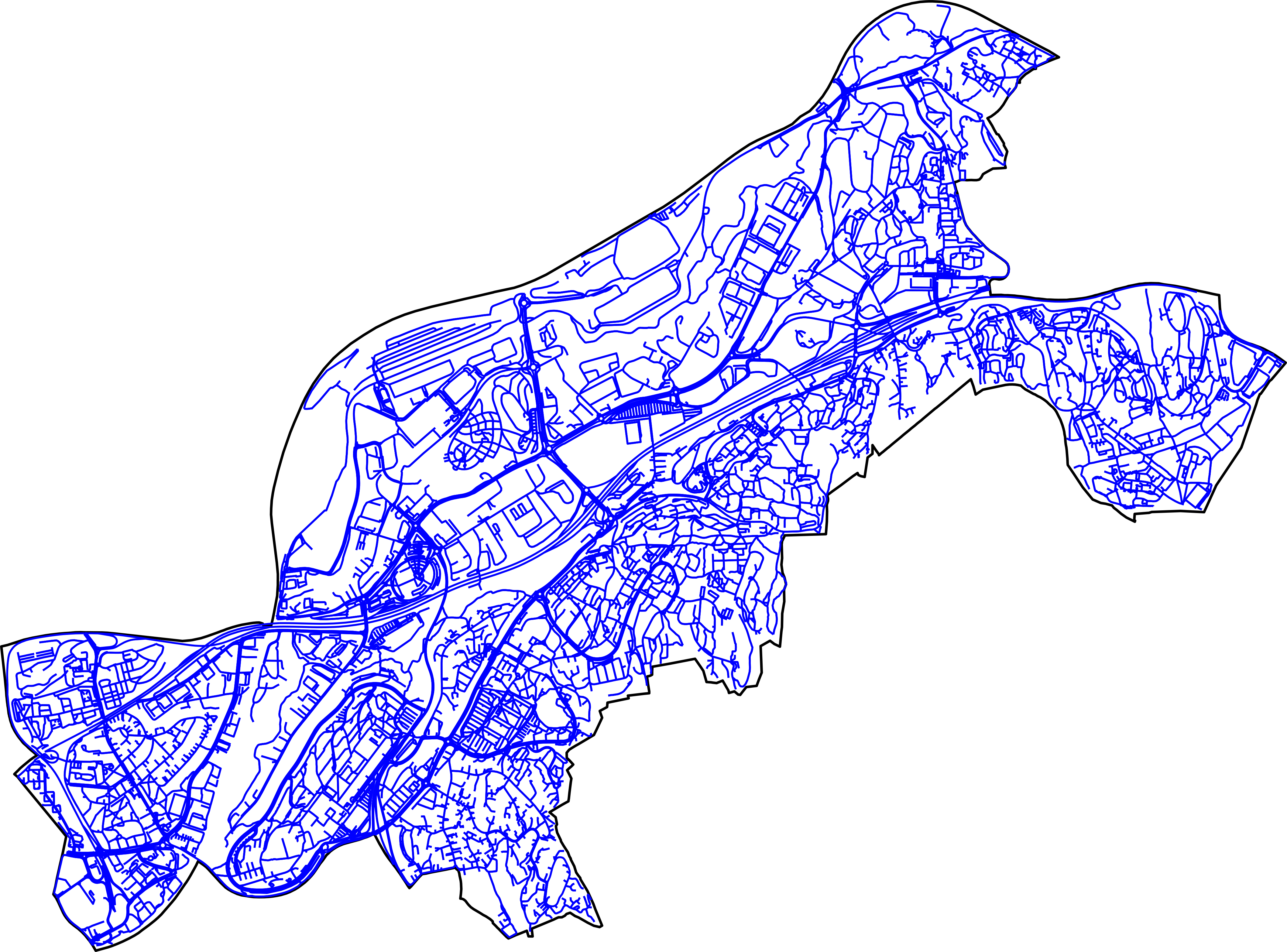}
        \caption{Retrieved road networks from OSM.}
        \label{fig:sub-1}
    \end{subfigure}%
    \hfill
    \begin{subfigure}[b]{0.45\textwidth}
        \centering
        \includegraphics[width=\linewidth]{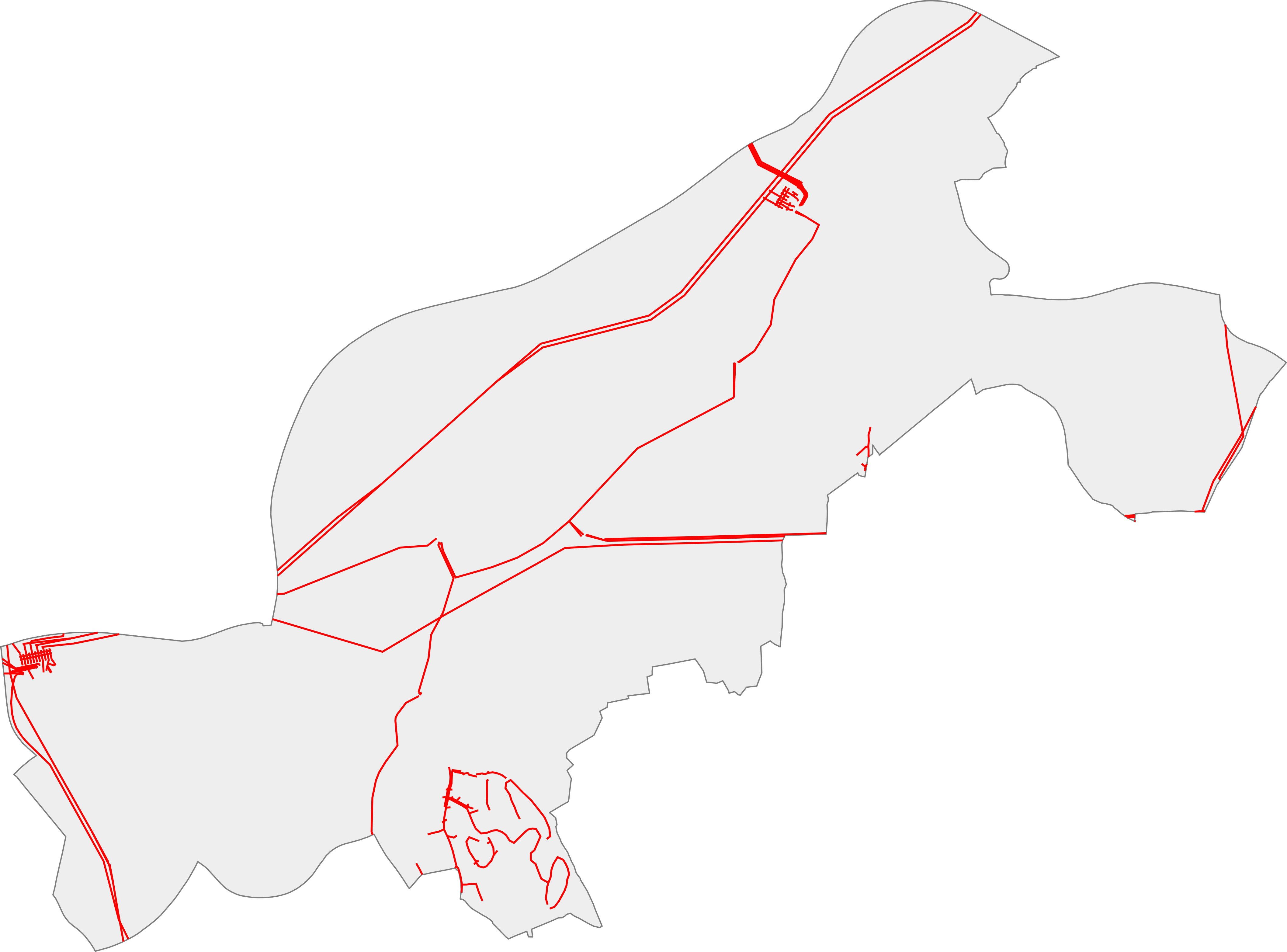}
        \caption{Retrieved power lines from OSM.}
        \label{fig:sub-2}
    \end{subfigure}
    \caption{An example of retrieved information (for the area of Alna, Oslo, Norway).}
    \label{fig:road_powerline}
\end{figure}

The API first downloads the road networks in the considered area from OpenStreetMap (OSM). We show how the road network is like in an example in Fig.~\ref{fig:sub-1}. We then discretize the road edges, and we interpolate the elevation for the road from the underlying COP30 DEM data. In this way, we render roads as 3D scatter paths. Similarly, the API retrieves power line geometries from OSM in the format of vectors, as shown in an example in Fig.~\ref{fig:sub-2}. Then these vectors undergo the same draping process, resulting in the visualization of energy corridors relative to the terrain.

\begin{figure}[tbhp]
    \centering
    \includegraphics[width=0.49\linewidth]{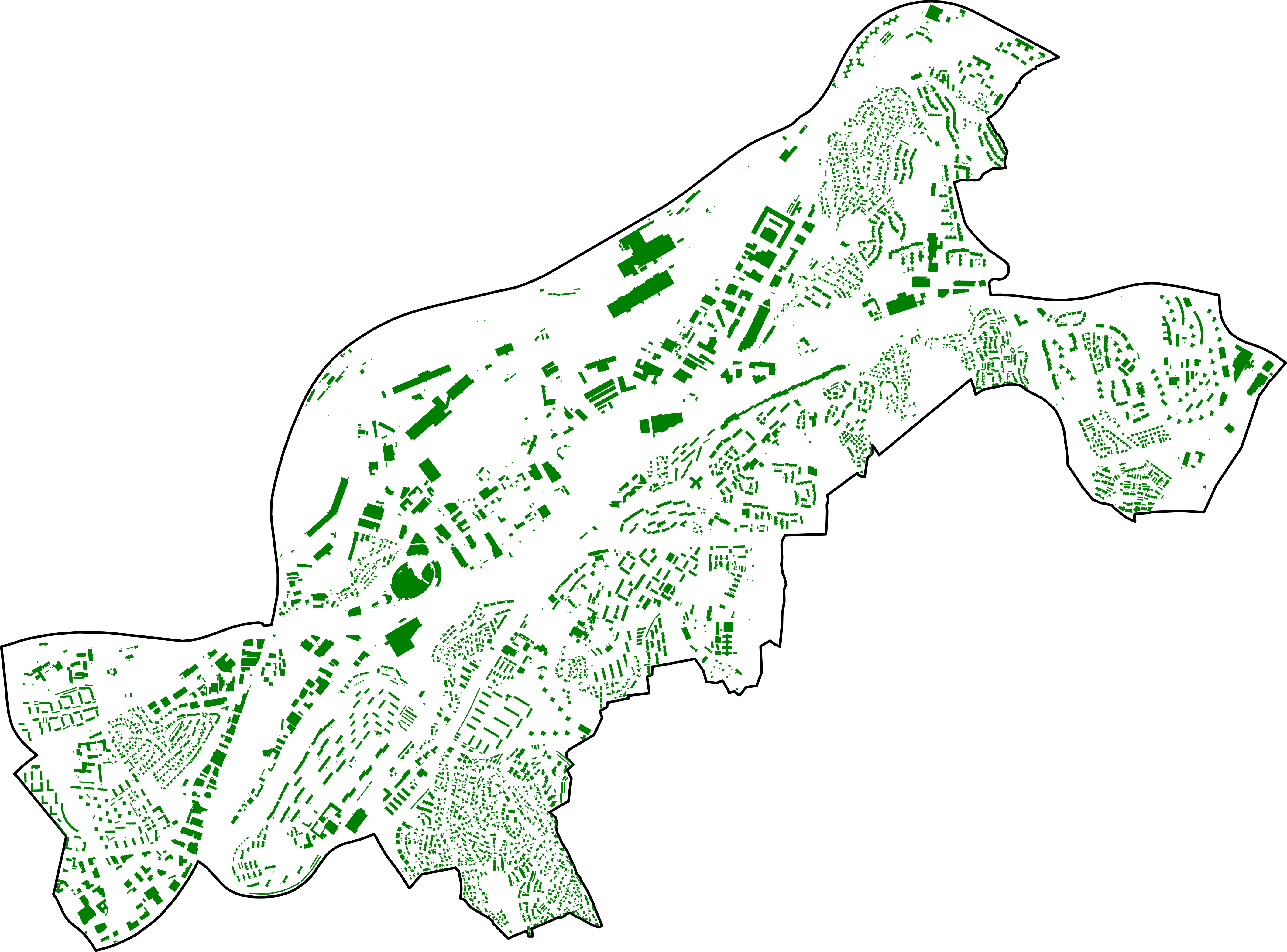}
    \caption{Retrieved building footprints from OSM (for the area of Alna, Oslo, Norway).}
    \label{fig:buildings}
\end{figure}

\subsection{Building extrusion}
\begin{figure}[tbhp]
    \centering
    \includegraphics[width=0.7\linewidth]{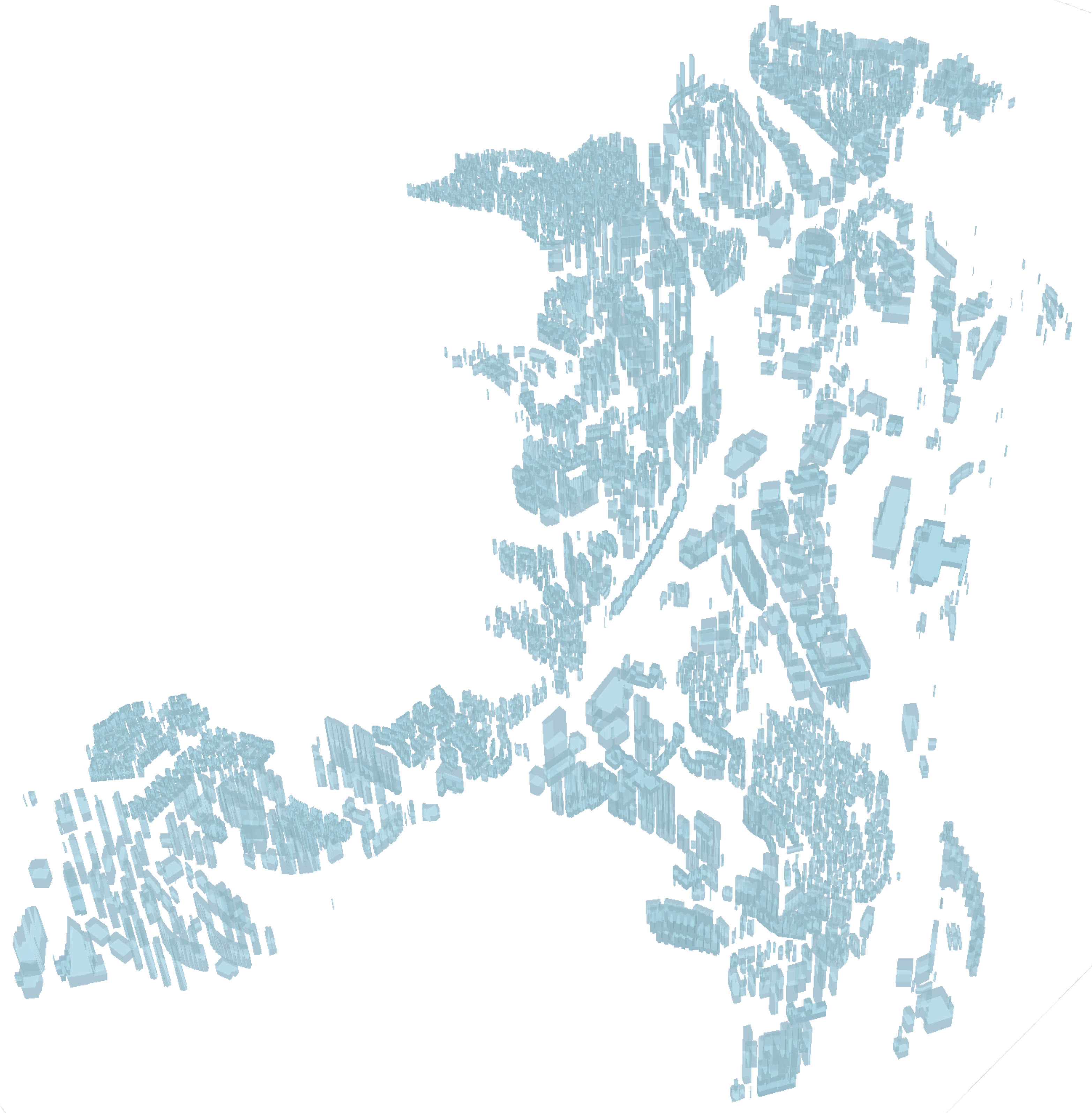}
    \caption{An example of extruded buildings (for the area of Alna, Oslo, Norway).}
    \label{fig:building_3d}
\end{figure}

Accurate building heights are often missing from standard OpenStreetMap data. We in this API solve this by integrating the data from \texttt{Overture Maps}. We start with fetching the building footprints in the considered area from OSM, as visualized in an example in Fig.~\ref{fig:buildings}. Then we query Overture Maps to gain the height of the OSM building footprints. After that, we create a base polygon at the terrain elevation of the area, and a roof polygon with the elevation added by the building height. We finally generate vertical faces to connect the base and roof, leading to a 3D mesh for each of the buildings, as shown in an example in Fig.~\ref{fig:building_3d}.

\section{Demonstration of how to use the API}

This section shows how to use our API via standard HTTP requests. The client-side interaction involves two steps: constructing the request URL and fetching the data, followed by rendering the resulting 3D model.

The API accepts parameters directly within the URL path. As shown in Fig.~\ref{lst1}, the request is formulated by concatenating the base URL with the necessary parameters: the \textit{OpenTopography} API key and the name of the target area. Note that this additional API key is free and can be obtained from the \textit{OpenTopography} platform\footnote{\url{https://opentopography.org}, accessed November 27, 2025.}.

The target area utilizes a specific, URL-friendly format (\textit{e.g.}, \texttt{Måøyna-Gulen-Vestland-Norway}) to define the precise area of interest. The system relies on the standard Python \texttt{requests} library to manage the \texttt{HTTP GET} transaction. The response body contains the 3D model, serialized as a complete \texttt{JSON} object describing the 3D representation.

\begin{figure}[tbhp] 
    \centering\caption{An example of constructing request to our API.}\label{lst1}
    \begin{lstlisting}[style=PythonStyle]
import requests
import json
from io import StringIO

api_key = '111222333444555666777888999'   
# Replace it by your OpenTopography API key, which is free from https://opentopography.org
target_place = 'Måøyna-Gulen-Vestland-Norway' 
# This is just an example of the place name, please replace the name in your case

# Below we show how to construct the url for the request to our API
BASE_URL = "https://cenergy3-qjbps.ondigitalocean.app"
api_request_url = f"{BASE_URL}/{api_key}/{target_place}"

# Fetch data from API by sending a request
try:
    response = requests.get(api_request_url)
    response.raise_for_status() # Check for the status of the request
    figure_dict = response.json()   # Gain the requested data in json format
except requests.exceptions.RequestException as e:
    print(f"Error fetching data: {e}")
    \end{lstlisting}
\end{figure}

The final step is to convert the retrieved \texttt{JSON} back into an interactive visualization object. As detailed in Fig.~\ref{lst:api_code_visualization}, the client-side code utilizes \texttt{plotly.graph\_objects.Figure} to deserialize the \texttt{JSON} object. This deserialization process instantly reconstructs the 3D scene. The use of \texttt{Plotly} guarantees that the output is immediately interactive, allowing the user to rotate, zoom, and inspect the contextual relationships between the energy infrastructure and the environment.

\begin{figure}[tbhp] 
    \centering\caption{An example of visualizing the requested data from our API}\label{lst:api_code_visualization}
    \begin{lstlisting}[style=PythonStyle]
import plotly.graph_objects as go

try:
    fig = go.Figure(figure_dict)
    fig.show()  # Display the interactive figure for the 3D model
except Exception as e:
    print(f"\nError in displaying the requested data: {e}")
    \end{lstlisting}
\end{figure}

\section{Interoperability with other libraries}

We refer interoperability of the developed API to the ability of different systems to communicate and exchange data effectively. This API achieves interoperability through the design detailed in the following.

\subsection{Standardized coordinates}
The API utilizes the \texttt{Web Mercator (EPSG:3857)} coordinate reference system for all 3D model generation. This standard, prevalent in web mapping (\textit{e.g.}, Google Maps, OpenStreetMap), ensures that the resulting 3D coordinates are readily compatible with most web-based visualization frameworks and geospatial data platforms without additional reprojections.

\subsection{Open data formats}
By relying on open data sources and utilizing standard file formats for intermediate steps, the API ensures that its internal processing is verifiable. The selection of \texttt{Overture Maps} and \texttt{OpenStreetMap} as primary data providers leverages a community-driven, non-proprietary data ecosystem that are scrutinized by the open communities.

\subsection{API output and integration}
The API's output is a \texttt{JSON} serialization of a \texttt{Plotly} Figure object. JSON is the lingua franca of web data exchange, making the model easily consumable by (i) web frontends, where \texttt{Plotly.js} allows the model to be natively embedded in JavaScript-based web applications, (ii) data science environments, where the \texttt{JSON} output can be instantly loaded and manipulated within Python, and (iii) cloud services, where we use the \texttt{FastAPI} framework to facilitate deployment on serverless or containerized platforms, making the 3D model generation accessible via standard \texttt{HTTP} requests. These design ensure the API is a modular, ``plug-and-play'' component in systems requiring on-demand 3D urban models.

\section*{Acknowledgment}

This work was supported by the PriTEM project funded by UiO:Energy Convergence Environments.

\bibliographystyle{unsrt}
\bibliography{references}

\begin{thebibliography}{10}

\bibitem{DBLP:journals/corr/abs-2510-09698}
Shiliang Zhang, Sabita Maharjan, Kai Strunz, and Jan~Christian Bryne.
\newblock Norwegian electricity in geographic dataset (noregeo).
\newblock {\em CoRR}, abs/2510.09698, 2025.

\bibitem{hersbach2019era5}
Hans Hersbach, Bill Bell, Paul Berrisford, Gionata Biavati, And{\'a}s Hor{\'a}nyi, J~Mu{\~n}oz~Sabater, Julien Nicolas, Carole Peubey, Raluca Radu, Iryna Rozum, et~al.
\newblock Era5 hourly data on single levels from 1979 to present.
\newblock {\em Copernicus Climate Change Service (C3S) Climate Data Store (CDS)}, 2019.

\bibitem{11174415}
Shiliang Zhang, Dyako Fatih, Fahmi Abdulqadir, Tobias Schwarz, and Xuehui Ma.
\newblock Extended vehicle energy dataset (eved): An enhanced large-scale dataset for vehicle energy consumption analysis.
\newblock In {\em 2025 IEEE 101st Vehicular Technology Conference (VTC2025-Spring)}, pages 01--07, 2025.

\bibitem{weinand2019spatial}
Jann~M Weinand, Russell McKenna, and Kai Mainzer.
\newblock Spatial high-resolution socio-energetic data for municipal energy system analyses.
\newblock {\em Scientific data}, 6(1):243, 2019.

\bibitem{9575223}
Shiliang Zhang.
\newblock An energy consumption model for electrical vehicle networks via extended federated-learning.
\newblock In {\em 2021 IEEE Intelligent Vehicles Symposium (IV)}, pages 354--361, 2021.

\bibitem{DBLP:journals/corr/abs-2203-08630}
Shiliang Zhang, Dyako Fatih, Fahmi Abdulqadir, Tobias Schwarz, and Xuehui Ma.
\newblock Extended vehicle energy dataset (eved): an enhanced large-scale dataset for deep learning on vehicle trip energy consumption.
\newblock {\em CoRR}, abs/2203.08630, 2022.

\bibitem{10538433}
Ruipeng Xu, Cuo Zhang, Daming Zhang, Zhao Yang~Dong, and Christine Yip.
\newblock Adaptive robust load restoration via coordinating distribution network reconfiguration and mobile energy storage.
\newblock {\em IEEE Transactions on Smart Grid}, 15(6):5485--5499, 2024.

\bibitem{DBLP:journals/corr/abs-2509-17095}
Jinbao Wang, Jun Liu, Shiliang Zhang, and Xuehui Ma.
\newblock Ultra-short-term solar power forecasting by deep learning and data reconstruction.
\newblock {\em CoRR}, abs/2509.17095, 2025.

\bibitem{oskar2025exploring}
Oskar Våle, Shiliang Zhang, Sabita Maharjan, and Gro Klæboe.
\newblock Exploring the interpretability of forecasting models for energy balancing market.
\newblock {\em Artificial Intelligence Science and Engineering}, 2025.

\bibitem{10738058}
Shiliang Zhang, Sabita Maharjan, Raul Shahi, and Xuehui Ma.
\newblock The impact of integration of renewable energy on imbalance settlement: Resilience analysis.
\newblock In {\em 2024 IEEE International Conference on Communications, Control, and Computing Technologies for Smart Grids (SmartGridComm)}, pages 98--104, 2024.

\bibitem{10738107}
Hui Zhang, Shiliang Zhang, Sabita Maharjan, and Yan Zhang.
\newblock P4s: Privacy-preserving personalized pricing scheme for smart grid.
\newblock In {\em 2024 IEEE International Conference on Communications, Control, and Computing Technologies for Smart Grids (SmartGridComm)}, pages 21--26, 2024.

\bibitem{zhang2020privacy}
Shiliang Zhang and Elad~M Schiller.
\newblock Privacy-preservation and the automotives.
\newblock In {\em 2020 AutoSPADA workshop, Gothenburg, Sweden}, 2020.

\bibitem{DBLP:journals/sensors/ZhangHSSA23}
Shiliang Zhang, Anton Hagermalm, Sanjin Slavnic, Elad~Michael Schiller, and Magnus Almgren.
\newblock Evaluation of open-source tools for differential privacy.
\newblock {\em Sensors}, 23(14):6509, 2023.

\bibitem{wylde2022cybersecurity}
Vinden Wylde, Nisha Rawindaran, John Lawrence, Rushil Balasubramanian, Edmond Prakash, Ambikesh Jayal, Imtiaz Khan, Chaminda Hewage, and Jon Platts.
\newblock Cybersecurity, data privacy and blockchain: A review.
\newblock {\em SN computer science}, 3(2):127, 2022.

\bibitem{DBLP:journals/corr/abs-2503-03539}
Shiliang Zhang, Sabita Maharjan, Lee~Andrew Bygrave, and Shui Yu.
\newblock Data sharing, privacy and security considerations in the energy sector: {A} review from technical landscape to regulatory specifications.
\newblock {\em CoRR}, abs/2503.03539, 2025.

\bibitem{DBLP:journals/corr/abs-2312-11564}
Daniel~Gerbi Duguma, Juliana Jia~Yu Zhang, Meysam Aboutalebi, Shiliang Zhang, Catherine Banet, Cato Bj{\o}rkli, Chinmayi~Prabhu Baramashetru, Frank Eliassen, Hui Zhang, Jonathan Muringani, Josef Noll, Knut~Inge Fostervold, Lars B{\"{o}}cker, Lee~Andrew Bygrave, Matin Bagherpour, Maunya~Doroudi Moghadam, Olaf Owe, Poushali Sengupta, Roman Vitenberg, Sabita Maharjan, Thiago Garrett, Yushuai Li, and Zhengyu Shan.
\newblock Privacy-preserving transactive energy systems: Key topics and open research challenges.
\newblock {\em CoRR}, abs/2312.11564, 2023.

\bibitem{SALVALAI2024114500}
Graziano Salvalai, Yunxi Zhu, and Marta {Maria Sesana}.
\newblock From building energy modeling to urban building energy modeling: A review of recent research trend and simulation tools.
\newblock {\em Energy and Buildings}, 319:114500, 2024.

\bibitem{el2024comprehensive}
Rasha~F El-Agamy, Hanaa~A Sayed, Arwa~M AL~Akhatatneh, Mansourah Aljohani, and Mostafa Elhosseini.
\newblock Comprehensive analysis of digital twins in smart cities: A 4200-paper bibliometric study.
\newblock {\em Artificial Intelligence Review}, 57(6):154, 2024.

\bibitem{9119753}
John~E. Vargas-Munoz, Shivangi Srivastava, Devis Tuia, and Alexandre~X. Falcão.
\newblock Openstreetmap: Challenges and opportunities in machine learning and remote sensing.
\newblock {\em IEEE Geoscience and Remote Sensing Magazine}, 9(1):184--199, 2021.

\end{thebibliography}

\end{document}